\documentclass[aip,jcp]{revtex4-1}
\usepackage{graphicx}
\usepackage{subfig}
\usepackage{mathptmx}      
\usepackage{latexsym}
\usepackage{gensymb}
\usepackage{hyperref}
\usepackage[version=4]{mhchem}
\usepackage{overpic}
\usepackage[textwidth=3.5cm, textsize=small]{todonotes}

\begin{document}


\title{Water/ice phase transition: the role of Zirconium Acetate, a compound with ice-shaping properties} 



\author{Moreno Marcellini}
\affiliation{Ceramic Synthesis and Functionalization Lab, UMR3080 CNRS/Saint-Gobain, 84306 Cavaillon, France}
\author{Francisco M. Fernandes}
\affiliation{Laboratoire de Chimie Matiere Condens\'ee de Paris, Universit\'e Pierre et Marie Curie, Sorbonne Universit\'es, UMR7574, 4 Place Jussieu, 75005 Paris, France}
\author{Sylvain Deville}
\email[]{sylvain.deville@saint-gobain.com}
\affiliation{Ceramic Synthesis and Functionalization Lab, UMR3080 CNRS/Saint-Gobain, 84306 Cavaillon, France}


\date{\today}

\begin{abstract}
Few compounds feature ice-shaping properties. The only compound reported to have ice-shaping properties similar to that of ice-shaping proteins, encountered in many organisms living at low temperature, is Zirconium Acetate.  When a Zirconium Acetate solution is frozen, oriented and perfectly hexagonal ice crystals can be formed and their growth follows the temperature gradient. To shed light on the water/ice phase transition while freezing Zirconium Acetate solution we carried out differential scanning calorimetry measurements. From our results, we estimate how many water molecules do not freeze because of their interaction with Zr cations. We estimate the colligative properties of the Zr concentration on the apparent critical temperature. We further show the phase transition is unaffected by the nature of the base which is used to adjust the pH. Our results provide thus new hints on the ice-shaping mechanism of Zirconium Acetate.
\end{abstract}


\maketitle 

\section{Introduction}
Water molecules, albeit composed of 3 atoms only, show complex behaviors and anomalies. Several theoretical models were proposed to describe its physico-chemical properties:~\footnote{See \href{http://www1.lsbu.ac.uk/water/water_models.html}{http://www1.lsbu.ac.uk/water/water\_models.html} } each model can suitably describe only certain characteristic. For example, the nucleation of ice in water is difficult to predict. Indeed, the water/ice phase transition is a very common but still puzzling and challenging phenomenon.

The nucleation of ice crystals occurs both in vapor and in liquid phases. Although it is difficult to maintain large volumes of water below 0\degree C without nucleation, a small volume can be hold in supercooled conditions.~\cite{Hobb:2010} Pradzynski \emph{et al.}~\cite{Prad:2012} found that at least 275$\pm$25 water molecules are required to initiate the ice nucleation. The freezing of water in distinct physico-chemical conditions results in different ice patterns, such as those of snow crystals, or the dendritic and cellular structure grown in supercooled water.~\cite{Shib:2003,Hobb:2010} 

The nucleation of water can be hindered by many solutes: \ce{NaCl} in sea water, for instance, lowers the freezing point by $\approx -1.8$\degree C. Colloidal particles can also depress the freezing point of water.~\cite{Pepp:2006} Ice is pure water. At the liquid-solid phase transition, water expels most of the ions and particles, which form brine channels between the ice crystals~\cite{Unte:1986} while releasing a latent heat $L=333.6$~kJ/kg.~\cite{Hobb:2010} Nevertheless, some chaotropic ions are easily incorporated in ice.

Zirconium Acetate (ZrAc), a moderately water soluble crystalline Zr source, decomposes to Zirconium Oxide on calcination at high temperature.~\cite{Tojo:1999} The physico-chemical processes leading to densification of liquid precursor solutions of ZrAc have been characterized by several methods.~\cite{Tosa:1994,Geic:1999,Geor:2012} The behavior of ZrAc solutions at temperature below the water/ice phase-transition was not investigated until ZrAc was used in ice-templating (also called freeze casting).~\cite{Devi:2008} Ice templating is a material processing route based on the growth of ice crystals in colloidal suspensions and the successive removal of water by sublimation. The porosity is thus templated by the ice crystals. The templated microstructures, architectures, and properties are related to the morphology of ice crystals grown during freezing. In this process, ZrAc can be used either as additive or as ceramic precursor. 

The use of ZrAc as an additive revealed its surprising ice-shaping properties.~\cite{Devi:2011} When a ZrAc solution, in a tight interval of pH (3.6-4.3) and concentration ($\approx$ 9 to 22.6 g/L of Zr), is frozen, faceted hexagonal ice crystals nucleate and grow along the direction of freezing. Such faceting is similar to the effects of ice-shaping proteins (ISPs)~\cite{Duma:1974} that in fishes,~\cite{Duma:1974} insects,~\cite{Knig:1986} and plants~\cite{Grif:1997} hinder, for example, the growth of large ice crystals and stabilize ice crystals in specific shapes. ZrAc (and the less powerful Zr hydroacetate)~\cite{Devi:2011,Mizr:2013} is so far the only inorganic compound that functions like ISPs. However, ISPs work at concentration of $\mu$mol or lower, whereas for ZrAc the lowest reported concentration for effective ice shaping properties is $\approx 0.10$~mol (9 g/L of Zr).~\cite{Devi:2011} The interactions of ISPs with ice are therefore probably much stronger than that of ZrAc.

To further understand the process that control the growth of ice crystals in ZrAc solutions, we froze several ZrAc solutions, prepared in two different ways, and measured by Differential Scanning Calorimetry (DSC) the liquid/solid phase transition. From our results, we assessed the amount of ice in the sample and estimated the radius of interaction of Zr cations. We further show that the phase transition is a colligative effect of the Zr concentration, independently of the base used for pH correction.

\section{Sample preparation}
A solution of ZrAc (in-house preparation of Saint-Gobain\footnote{The ice-shaping properties of ZrAc are independent of the supplier.}) with initial Zr concentration of 22.6~g/l (solution \#1) and original pH~$=2.6$, was diluted in deionized water to obtain two other solutions with equivalent starting Zr concentrations of 13.3~g/l (pH~$=3.2$, solution \#3), 18~g/l (pH~$=3.1$, solution \#2): each was prepared in two batches. The pH of one of the batches was then adjusted to pH~$=4.0\pm 0.1$ by adding 25 mol/L NaOH solution and HCl solution at 37\%~wt (Sigma Aldrich): solutions \#\emph{X}(NaOH). The pH of the latter batch was carefully set, without overshooting, to pH~$\approx 4$ with TMA-OH solution (tetramethylammonium hydroxide, Sigma Aldrich, concentration of 10~wt.\% in water): solutions \#\emph{X}(TMAOH). 
\begin{table}[h]
\begin{center}
\caption{Samples and estimated Zr concentration}
\label{tab:samples} 
\begin{tabular}{cccc}
\hline
Sample & NaOH/HCl (g/L) & TMA-OH (g/L) & Before pH adj.\\
\hline
\#1 & $\approx$~22.6 & $\approx$~16.7& 22.6 \\
\#2 & $\approx$~17.7 & $\approx$~12.0& 18 \\
\#3 & $\approx$~13.1 & $\approx$~11.2& 13.3 \\
\hline
\end{tabular}
\end{center}
\end{table}
TMA-OH is a strong chaotrope hydrophylic molecule believed to weaken the hydrogen bonds network of water. It can entrap several molecules of water by hydrogen bonds~\cite{Koga:2011,Nils:2016} producing chlatrates.~\cite{Moot:1990} Both \ce{Na+} and \ce{Cl-} are in the middle of the the Hofmeister Series~\cite{Zhan:2006,Coll:2009,Zavi:2016} whereas TMA-OH is the strongest chaotrope. The acetate group is therefore more kosmotrope than \ce{Cl-}.~\cite{LoN:2012} By probing two different sets of solutions, we can also assess whether the chaotrope/kosmotrope natures of the two bases affect the liquid/solid phase transition. The low initial concentration of the TMA-OH solution and the its high pK$_b$=4.2 caused large shifts in the final Zr concentration after the pH adjustment. The final concentration of the solutions adjusted with NaOH/HCl was left approximately unchanged, see Table~\ref{tab:samples}. For each solution, we also prepared the corresponding reference solutions made up by a volume of water equivalent to the initial volume of the three solutions and the same amount of bases needed to adjust the pHs.

DSC measurements were carried out in N$_2$ atmosphere. The temperature rate was set to 5\degree C/min in the close cycle from +40 to -80\degree C. We recorded two cycles per solution. The chosen freezing rate corresponds to ice shaping conditions as observed by freeze casting.   The data analysis was performed with the Universal Analysis 2000 (TA Instruments) software. 

\section{Results \& Discussion}
\begin{figure}[h]
\includegraphics[angle=-0,width=1.0\linewidth]{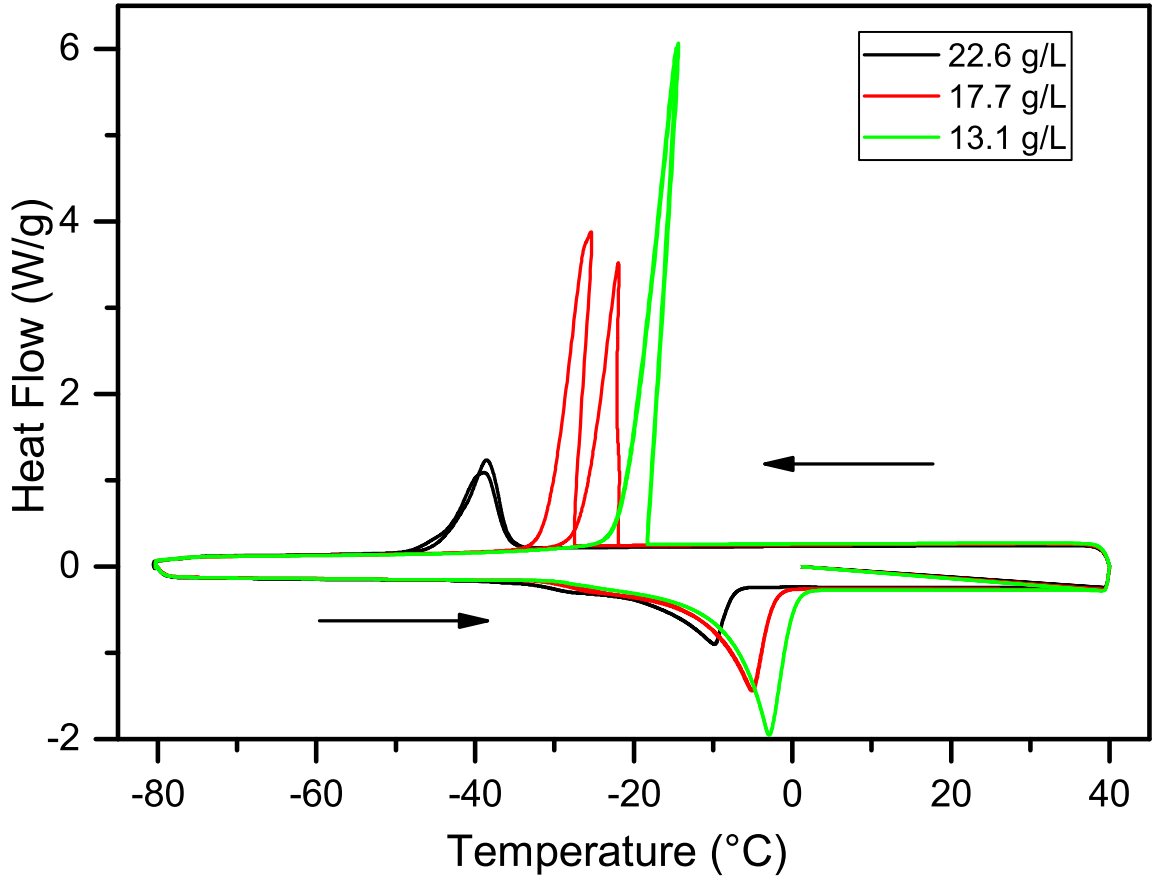}
\caption{Full thermograms of \#\emph{X}(NaOH) solutions. Arrows indicate the direction of the cycle.\iffalse The heating cycles are used to compute the actual melting (freezing) temperatures and enthalpies \fi} 
\label{fig:NaOH}
\end{figure}

\begin{figure}[h]
\includegraphics[angle=-0,width=1.0\linewidth]{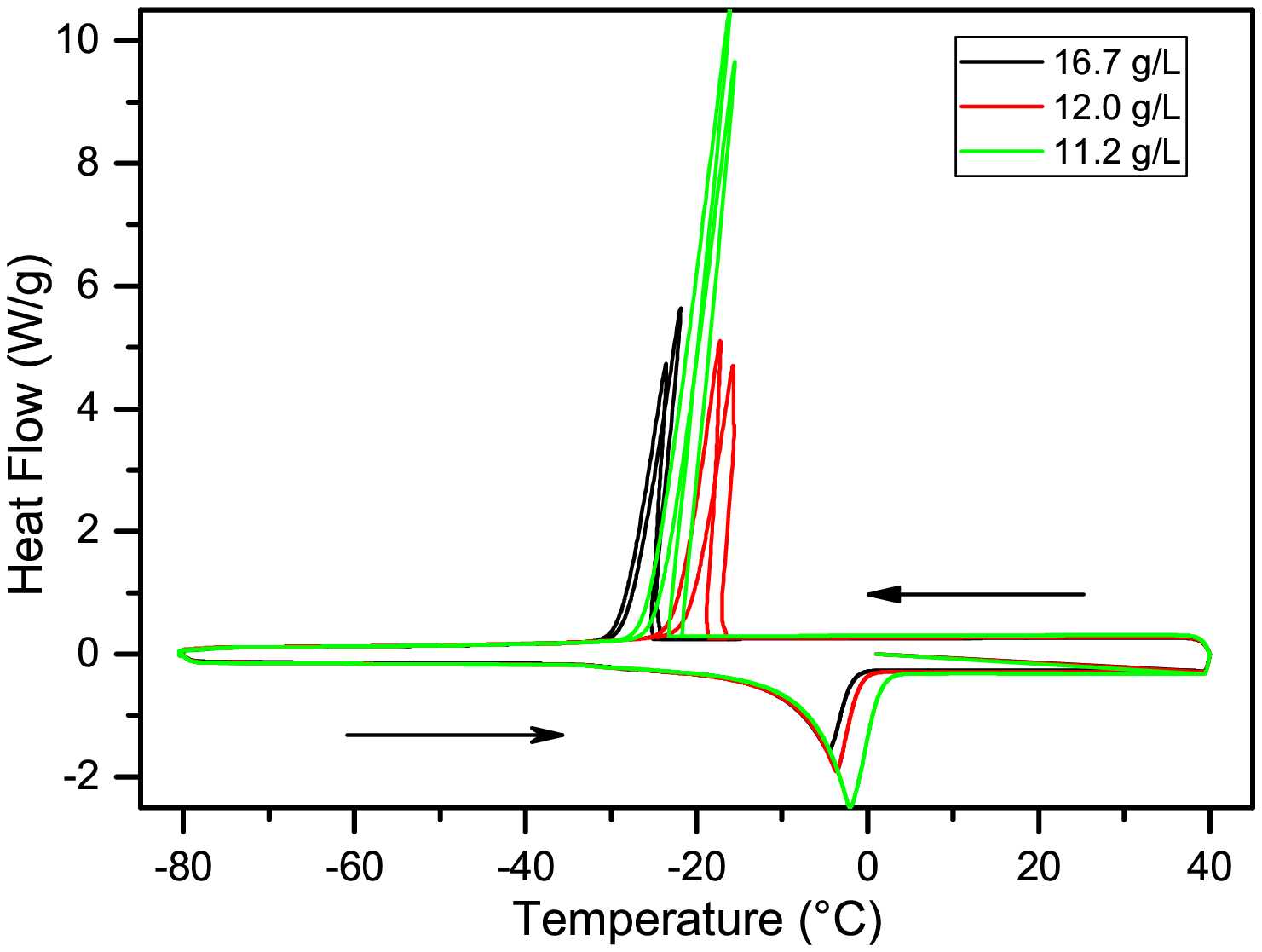}
\caption{Like in Fig.~\ref{fig:NaOH}, thermograms for the \#\emph{X}(TMAOH) solutions. Similarly to the \#\emph{X}(NaOH) case, the apparent freezing temperatures correlate with the Zr concentration.}
\label{fig:TMAOH}
\end{figure}

\begin{figure}[h]
\includegraphics[angle=-0,width=1.0\linewidth]{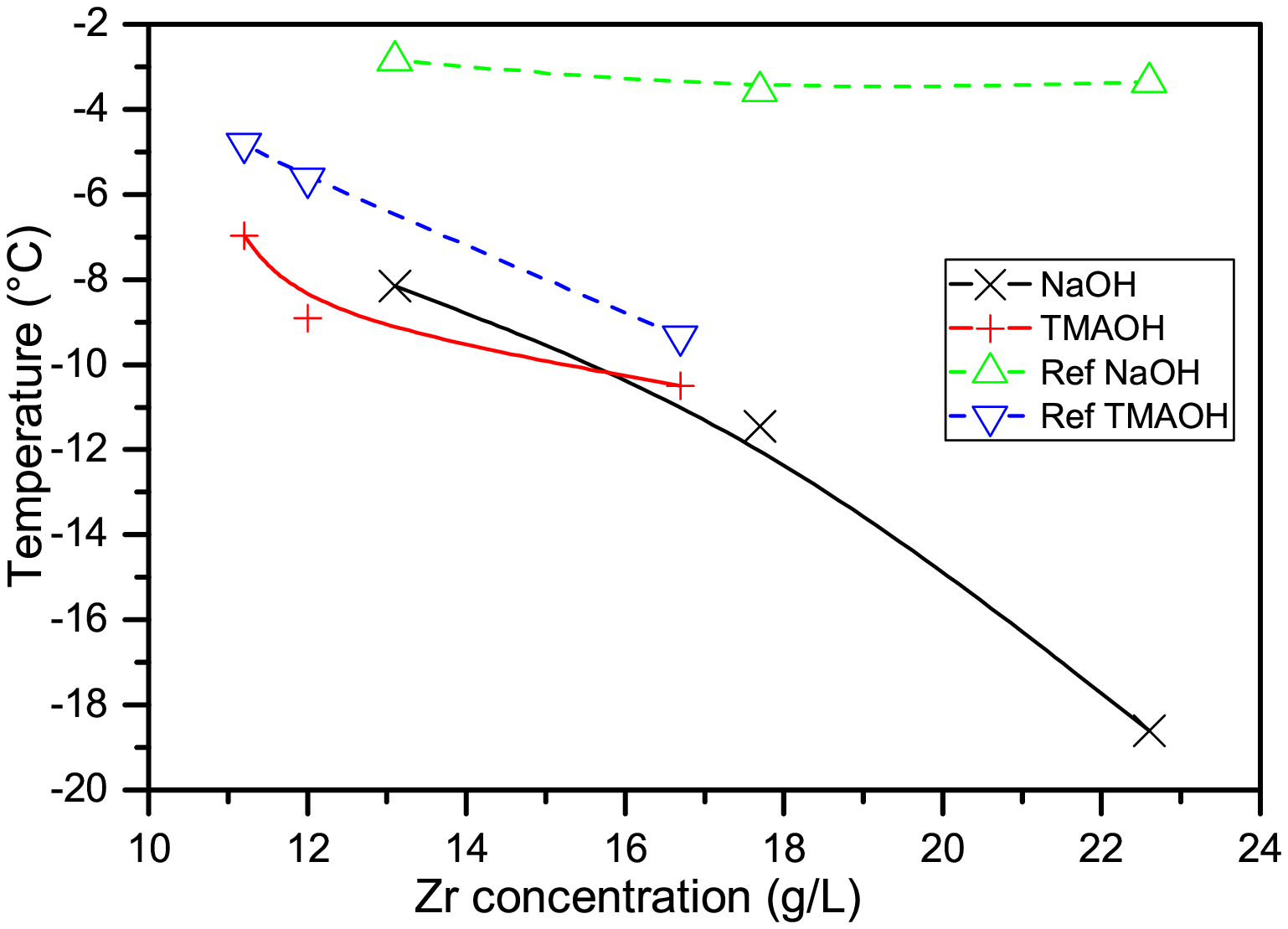}
\caption{$T_c$ of the solutions. $T_c$ is a colligative property of Zr concentration. (The smoothed lines are guide for the eyes.)} 
\label{fig:Tdrop}
\end{figure}

The thermograms for the \#\emph{X}(NaOH) solutions are shown in Fig.~\ref{fig:NaOH}. During the cooling semi-cycle, the freezing (critical) point is apparently depressed to temperatures $\leq -20$~\degree C, because the phase transition occurs in supercooled state, triggering an avalanche-alike dendritic formation of ice crystals.~\cite{Shib:2003} The temperature increases during the phase transition for \#3(NaOH). This  artifact is due to the instrumentation not able to fully withdraw the latent heat. Likewise, the double peak for the \#2(NaOH) is due to the unpredictable ice nucleation in supercooled conditions. Similar experiments on the \#\emph{X}(TMAOH) solutions give the results shown in Fig.~\ref{fig:TMAOH}.
Hereby we define the apparent critical temperature $T_c$ as the on-set temperature of the solid/liquid phase transition: these values are reported in Tab.~\ref{tbl:fusion} and plotted in Fig.~\ref{fig:Tdrop} with the relative results of reference solutions.
\begin{table}[h]
\begin{center}
\caption{Apparent \iffalse critical temperatures\fi $T_c$ of ZrAc and reference solutions. (The latter in parentheses.)}
\label{tbl:fusion} 
\begin{tabular}{c|cc|cc}
\hline
Sample & Zr (g/L) & NaOH/HCl (\degree C) & Zr (g/L) &TMA-OH (\degree C) \\
\hline
\#1 & 22.6 & -18.61 (-3.37) & 16.7 &  -10.50 (-9.34)\\
\#2 & 17.7 &  -11.45 (-3.58) &  12.0 &  -8.91 (-5.62)\\
\#3 & 13.1 & -8.15 (-2.86) &  11.2 &  -6.97 (-4.80)\\
\hline
\end{tabular}
\end{center}
\end{table}
The results in Tab.~\ref{tbl:fusion} tell us that the apparent $T_c$ is a colligative property of ZrAc concentration. For example, \#2(NaOH) and \#3(NaOH) share close concentrations with \#1(TMAOH) and \#2(TMAOH) respectively, and their $T_c$ are quite similar. When the pHs of the original solutions are adjusted with NaOH/HCl, the large difference between the $T_c$ of the samples and reference solutions hints to how strong the interaction between the solute (ZrAc) and the surrounding water shell is. ZrAc in water solution hydrolizes and self-assembles in tetrameric basic units,~\cite{Mak:1968,Clea:1990} where the Zr atoms are connected by hydroxo-bridges,~\cite{Hagf:2004} whose stability was confirmed by computation.~\cite{Rao:2007} In acidic conditions the reaction

tetramers \ce{->} octamers~\cite{Sing:1996} \ce{->} stacks~\cite{Goss:2014}

\noindent occurs depending on the concentration, the pH, and the time. We believe  the tetramers organize in stacks with bound acetate groups, due to the strong interaction of Zr with carboxylic group.~\cite{Clea:1964} The length of such stacks should be inversely proportional to the Zr concentration.~\cite{Goss:2014} Such structure should resemble the one of ISPs from, for example, spruce budworm~\cite{Lein:2002} or from winter flounder.~\cite{Sun:2014} This large structures can interact with several water shells thanks to the ability of acetate groups to alter the hydrogen bonds network in the neighborhood. We hypothesize that likewise ISPs, the ZrAc stacks should adsorb onto the blooming ice crystals and causes their directional and faceted growth.   

\begin{figure}[h]
\includegraphics[angle=-0,width=1.0\linewidth]{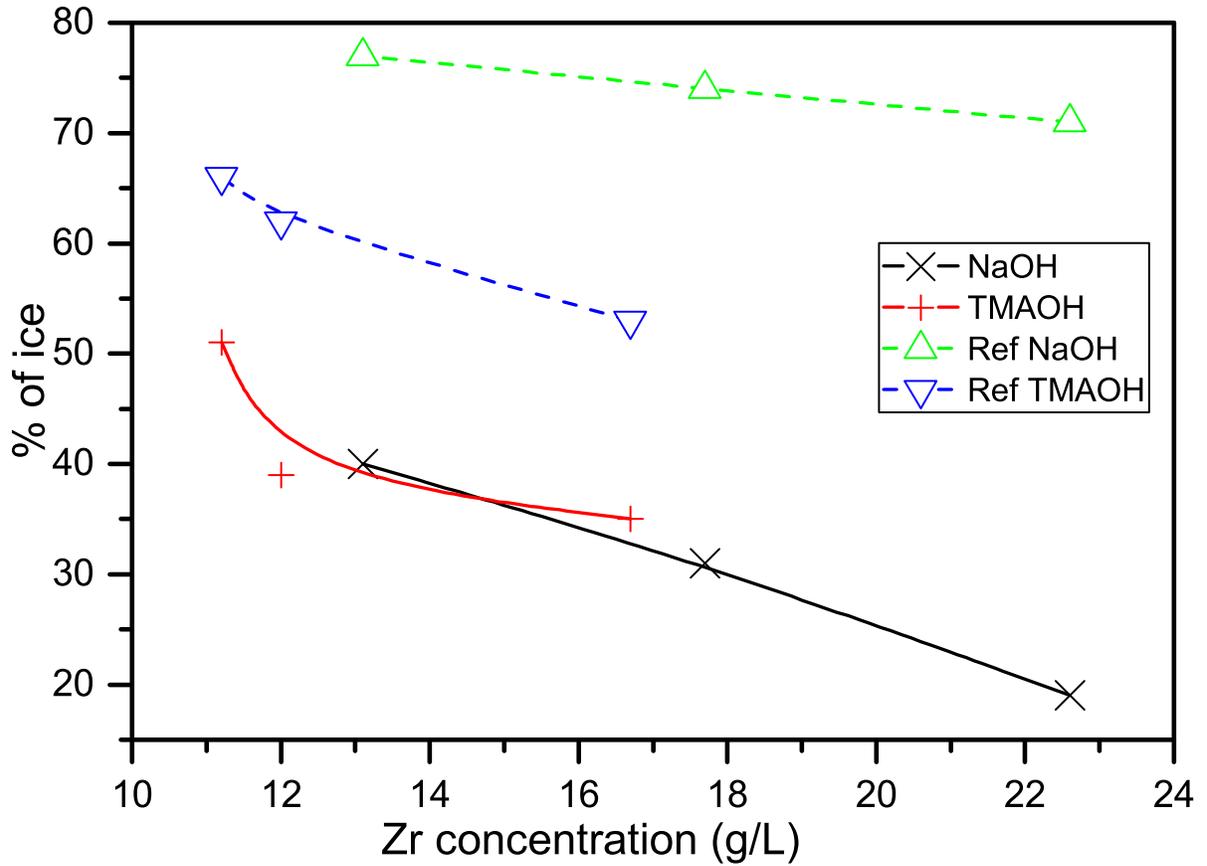}
\caption{Ice contents in the solutions. (The smoothed lines are guides for the eyes.)} 
\label{fig:Ice}
\end{figure}

The latent heat $L$ released at the phase transition depends only on the amount of water that participates to the phase transition. The ratio between the experimental and the tabulated value~\cite{Hobb:2010} represents the amount of water that has taken part to the phase transition. These results, computed on the heating semi-cycle, are in Tab.~\ref{tbl:percent} and plotted in Fig.~\ref{fig:Ice}.
\begin{table}[h]
\begin{center}
\caption{Amount (percent) of frozen water in each sample. In parentheses the values for the reference solutions.}
\label{tbl:percent} 
\begin{tabular}{c|cc|cc}
\hline
Sample & Zr (g/L) & NaOH/HCl (\%) & Zr (g/L) & TMA-OH (\%) \\
\hline
\#1 & 22.6 & 19 (71) & 16.7 & 35 (53) \\
\#2 & 17.7 & 31 (74) & 12.0 & 39 (62) \\
\#3 & 13.1 & 40 (77) & 11.2 & 51 (66) \\
\hline
\end{tabular}
\end{center}
\end{table}
Clearly the physico-chemical properties of the bases do not interfere with the phase transition. For example, the samples \#2(NAOH) and \#1(TMAOH) contain similar amount of ZrAc (17.7 g/L versus 16.7 g/L) and the ice content is similar. Thus, we may conclude the thermodynamic of the process is due exclusively to ZrAc, strengthening the hypothesis that the ice-shaping properties of ZrAc are colligative.
If we approximate the solutions as composed by water and ZrAc, we can estimate, from the amount of ice, how many molecules of water do not participate to the phase transition due to their interaction with the Zr cations. This approach is eased in the case of \#\emph{X}(NaOH) solutions, where the \ce{NaOH/HCl} content is minimal. Thus, for the solution \#1(NaOH), we estimate that 180 water molecules per Zr cation (or 720 per Zr tetramer) do not contribute to the phase transition.  We can represent it as a sphere of radius $r\approx 1.74$~nm ($r\approx 2.8$~nm for the tetramer) (see Tab.~\ref{tbl:radii}). To give an idea of how large the sphere of interaction for Zr cations is, the  Zr-Zr interdistance is $\approx 0.32$~nm. For example, the hydrophylic TMA-OH binds 25 water molecules by hydrogen bond.~\cite{Koga:2011,Koga:2013}
\begin{table}[h]
\begin{center}
\begin{tabular}{cccc}
\hline
Sample & \ce{H2O} (mol) & radius (nm) & Zr~$\rho_f$~(g/L) \\
\hline
\#1(NAOH) & 180 & 1.74 & 27.9 \\
\#2(NAOH) & 198 & 1.80 & 25.7\\
\#3(NAOH) & 232 & 1.89 & 21.8 \\
\hline
\end{tabular}
\end{center}
\caption{Water molecules per Zr cation; Radii of interaction per Zr cation; Zr concentration after phase transition in water.}\label{tbl:radii} 
\end{table}
The percent of frozen water in Table~\ref{tbl:percent} represents, to a first approximation, the volume of ice. Since the ice turns into porosity in the final material, this volume is thus an indirect measurement of the overall porosity in ice-templated materials. 
\begin{figure}[h]
\includegraphics[angle=-0,width=1.0\linewidth]{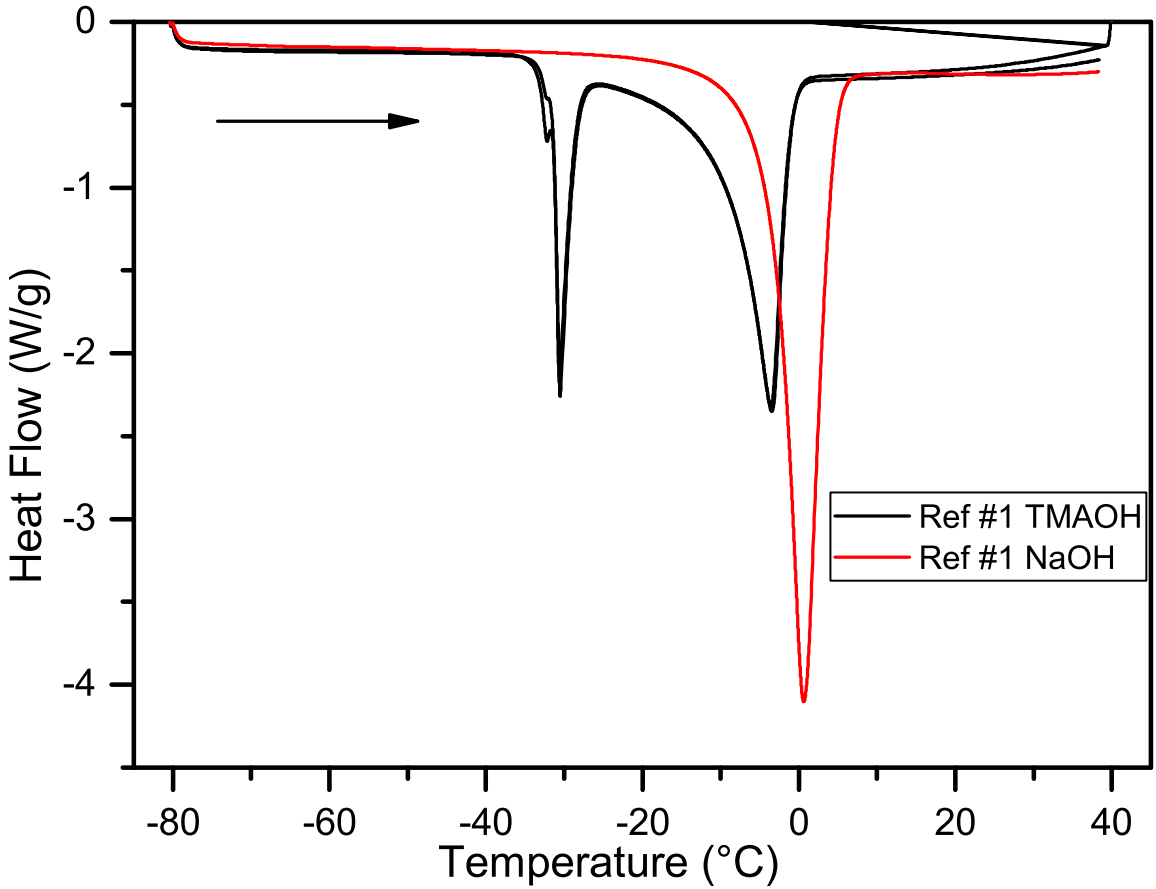}
\caption{DSC thermogram of the solid/liquid phase transitions for two reference solutions. \iffalse Arrow indicates the direction of the cycle.\fi Note the double melting peak for the solution loaded with TMA-OH.}
\label{fig:BothRef}
\end{figure}
While measuring the reference solutions we noted a secondary peak at lower temperature occurring for the solution loaded exclusively with TMA-OH, Fig.~\ref{fig:BothRef}. We believe this extra phase transition could result from the formation of TMA-OH clathrates.~\cite{Moot:1990}

\section{Conclusions}
The freezing point $T_c$ of the liquid/solid phase transition of water depends on the ionic content. We have shown that $T_c$ is a colligative properties of the ionic content of ZrAc, probably the only inorganic compound able to mimic the effects of ISPs. Further, the amount of ice  depends on the concentration of Zr in solution, independently on how the pH of the solutions is stabilized. The chaotrope hydrophylic TMA-OH has a very strong influence on the freezing process. The clathrate organization of TMA-OH is destroyed by the present of ZrAc. We have estimated the sphere of interaction of the Zr cations on water molecules. The Zr cations display a radius of interaction of few nm, that comprises hundreds of water molecules. These results give an hint on the chaotropic and hydrophylic strength of Zr either tetramers or stacks, which should populate the solutions. In the 3 cases investigated here, the number of water molecules that do not nucleate is different. This reinforces the hypothesis the the length of Zr tetramers stack is inversely proportional to the concentration. The final concentration of Zr in the remnant liquid water should, thus, not be the same. Nevertheless, we can not exclude further effects due to the pH of the studied solutions. We think spectroscopic and microscopic measurements could be used to better understand the ice-shaping mechanism of ZrAc.   

\section*{Funding}
The research leading to these results has received funding from the European Research Council under the European Community's Seventh Framework Programme (FP7/2007-2013) Grant Agreement no. 278004, \emph{FreeCo}. 

\begin{acknowledgments}
We acknowledge Dr. C. Noirjean and D. Dedovets for their comments on the manuscript.
\end{acknowledgments}

\bibliography{papero}

\providecommand{\noopsort}[1]{}\providecommand{\singleletter}[1]{#1}%
\begin{thebibliography}{34}%
\makeatletter
\providecommand \@ifxundefined [1]{%
 \@ifx{#1\undefined}
}%
\providecommand \@ifnum [1]{%
 \ifnum #1\expandafter \@firstoftwo
 \else \expandafter \@secondoftwo
 \fi
}%
\providecommand \@ifx [1]{%
 \ifx #1\expandafter \@firstoftwo
 \else \expandafter \@secondoftwo
 \fi
}%
\providecommand \natexlab [1]{#1}%
\providecommand \enquote  [1]{``#1''}%
\providecommand \bibnamefont  [1]{#1}%
\providecommand \bibfnamefont [1]{#1}%
\providecommand \citenamefont [1]{#1}%
\providecommand \href@noop [0]{\@secondoftwo}%
\providecommand \href [0]{\begingroup \@sanitize@url \@href}%
\providecommand \@href[1]{\@@startlink{#1}\@@href}%
\providecommand \@@href[1]{\endgroup#1\@@endlink}%
\providecommand \@sanitize@url [0]{\catcode `\\12\catcode `\$12\catcode
  `\&12\catcode `\#12\catcode `\^12\catcode `\_12\catcode `\%12\relax}%
\providecommand \@@startlink[1]{}%
\providecommand \@@endlink[0]{}%
\providecommand \url  [0]{\begingroup\@sanitize@url \@url }%
\providecommand \@url [1]{\endgroup\@href {#1}{\urlprefix }}%
\providecommand \urlprefix  [0]{URL }%
\providecommand \Eprint [0]{\href }%
\providecommand \doibase [0]{http://dx.doi.org/}%
\providecommand \selectlanguage [0]{\@gobble}%
\providecommand \bibinfo  [0]{\@secondoftwo}%
\providecommand \bibfield  [0]{\@secondoftwo}%
\providecommand \translation [1]{[#1]}%
\providecommand \BibitemOpen [0]{}%
\providecommand \bibitemStop [0]{}%
\providecommand \bibitemNoStop [0]{.\EOS\space}%
\providecommand \EOS [0]{\spacefactor3000\relax}%
\providecommand \BibitemShut  [1]{\csname bibitem#1\endcsname}%
\let\auto@bib@innerbib\@empty
\bibitem [{Note1()}]{Note1}%
  \BibitemOpen
  \bibinfo {note} {See \protect \href
  {http://www1.lsbu.ac.uk/water/water_models.html}{http://www1.lsbu.ac.uk/water/water\protect
  \_models.html}}\BibitemShut {NoStop}%
\bibitem [{\citenamefont {Hobbs}(2010)}]{Hobb:2010}%
  \BibitemOpen
  \bibfield  {author} {\bibinfo {author} {\bibfnamefont {P.~V.}\ \bibnamefont
  {Hobbs}},\ }\href@noop {} {\emph {\bibinfo {title} {{Ice Physics}}}},\ Oxford
  Classic Texts in the Physical Sciences\ (\bibinfo  {publisher} {Oxford
  University Press},\ \bibinfo {year} {2010})\BibitemShut {NoStop}%
\bibitem [{\citenamefont {Pradzynski}\ \emph {et~al.}(2012)\citenamefont
  {Pradzynski}, \citenamefont {Forck}, \citenamefont {Zeuch}, \citenamefont
  {Slav{\'\i}{\v c}ek},\ and\ \citenamefont {Buck}}]{Prad:2012}%
  \BibitemOpen
  \bibfield  {author} {\bibinfo {author} {\bibfnamefont {C.~C.}\ \bibnamefont
  {Pradzynski}}, \bibinfo {author} {\bibfnamefont {R.~M.}\ \bibnamefont
  {Forck}}, \bibinfo {author} {\bibfnamefont {T.}~\bibnamefont {Zeuch}},
  \bibinfo {author} {\bibfnamefont {P.}~\bibnamefont {Slav{\'\i}{\v c}ek}}, \
  and\ \bibinfo {author} {\bibfnamefont {U.}~\bibnamefont {Buck}},\ }\href
  {\doibase 10.1126/science.1225468} {\bibfield  {journal} {\bibinfo  {journal}
  {Science}\ }\textbf {\bibinfo {volume} {337}},\ \bibinfo {pages} {1529}
  (\bibinfo {year} {2012})}\BibitemShut {NoStop}%
\bibitem [{\citenamefont {Shibkov}\ \emph {et~al.}(2003)\citenamefont
  {Shibkov}, \citenamefont {Golovin}, \citenamefont {Zheltov}, \citenamefont
  {Korolev},\ and\ \citenamefont {Leonov}}]{Shib:2003}%
  \BibitemOpen
  \bibfield  {author} {\bibinfo {author} {\bibfnamefont {A.}~\bibnamefont
  {Shibkov}}, \bibinfo {author} {\bibfnamefont {Y.}~\bibnamefont {Golovin}},
  \bibinfo {author} {\bibfnamefont {M.}~\bibnamefont {Zheltov}}, \bibinfo
  {author} {\bibfnamefont {A.}~\bibnamefont {Korolev}}, \ and\ \bibinfo
  {author} {\bibfnamefont {A.}~\bibnamefont {Leonov}},\ }\href
  {http://www.sciencedirect.com/science/article/pii/S0378437102015170}
  {\bibfield  {journal} {\bibinfo  {journal} {Physica A: Statistical Mechanics
  and its Applications}\ }\textbf {\bibinfo {volume} {319}},\ \bibinfo {pages}
  {65} (\bibinfo {year} {2003})}\BibitemShut {NoStop}%
\bibitem [{\citenamefont {Peppin}, \citenamefont {Elliott},\ and\ \citenamefont
  {Worster}(2006)}]{Pepp:2006}%
  \BibitemOpen
  \bibfield  {author} {\bibinfo {author} {\bibfnamefont {S.~S.}\ \bibnamefont
  {Peppin}}, \bibinfo {author} {\bibfnamefont {J.~A.~W.}\ \bibnamefont
  {Elliott}}, \ and\ \bibinfo {author} {\bibfnamefont {M.~G.}\ \bibnamefont
  {Worster}},\ }\href {\doibase 10.1017/S0022112006009268} {\bibfield
  {journal} {\bibinfo  {journal} {Journal of Fluid Mechanics}\ }\textbf
  {\bibinfo {volume} {554}},\ \bibinfo {pages} {147} (\bibinfo {year}
  {2006})}\BibitemShut {NoStop}%
\bibitem [{\citenamefont {Untersteiner}(1986)}]{Unte:1986}%
  \BibitemOpen
  \bibinfo {editor} {\bibfnamefont {N.}~\bibnamefont {Untersteiner}},\ ed.,\
  \href {\doibase 10.1007/978-1-4899-5352-0} {\emph {\bibinfo {title} {{The
  Geophysics of Sea Ice}}}},\ \bibinfo {edition} {1st}\ ed.\ (\bibinfo
  {publisher} {Springer US},\ \bibinfo {year} {1986})\BibitemShut {NoStop}%
\bibitem [{\citenamefont {Tojo}\ \emph {et~al.}(1999)\citenamefont {Tojo},
  \citenamefont {Atake}, \citenamefont {Mori},\ and\ \citenamefont
  {Yamamura}}]{Tojo:1999}%
  \BibitemOpen
  \bibfield  {author} {\bibinfo {author} {\bibfnamefont {T.}~\bibnamefont
  {Tojo}}, \bibinfo {author} {\bibfnamefont {T.}~\bibnamefont {Atake}},
  \bibinfo {author} {\bibfnamefont {T.}~\bibnamefont {Mori}}, \ and\ \bibinfo
  {author} {\bibfnamefont {H.}~\bibnamefont {Yamamura}},\ }\href
  {http://dx.doi.org/10.1023/A:1010159807127} {\bibfield  {journal} {\bibinfo
  {journal} {Journal of Thermal Analysis and Calorimetry}\ }\textbf {\bibinfo
  {volume} {57}},\ \bibinfo {pages} {447} (\bibinfo {year} {1999})}\BibitemShut
  {NoStop}%
\bibitem [{\citenamefont {Tosan}\ \emph {et~al.}(1994)\citenamefont {Tosan},
  \citenamefont {Durand}, \citenamefont {Roubin}, \citenamefont {Chassagneux},\
  and\ \citenamefont {Bertin}}]{Tosa:1994}%
  \BibitemOpen
  \bibfield  {author} {\bibinfo {author} {\bibfnamefont {J.-L.}\ \bibnamefont
  {Tosan}}, \bibinfo {author} {\bibfnamefont {B.}~\bibnamefont {Durand}},
  \bibinfo {author} {\bibfnamefont {M.}~\bibnamefont {Roubin}}, \bibinfo
  {author} {\bibfnamefont {F.}~\bibnamefont {Chassagneux}}, \ and\ \bibinfo
  {author} {\bibfnamefont {F.}~\bibnamefont {Bertin}},\ }\href {\doibase
  http://dx.doi.org/10.1016/0022-3093(94)90116-3} {\bibfield  {journal}
  {\bibinfo  {journal} {Journal of Non-Crystalline Solids}\ }\textbf {\bibinfo
  {volume} {168}},\ \bibinfo {pages} {23 } (\bibinfo {year}
  {1994})}\BibitemShut {NoStop}%
\bibitem [{\citenamefont {Geiculescu}\ and\ \citenamefont
  {Spencer}(1999)}]{Geic:1999}%
  \BibitemOpen
  \bibfield  {author} {\bibinfo {author} {\bibfnamefont {A.~C.}\ \bibnamefont
  {Geiculescu}}\ and\ \bibinfo {author} {\bibfnamefont {H.~G.}\ \bibnamefont
  {Spencer}},\ }\href {\doibase 10.1023/A:1008769204022} {\bibfield  {journal}
  {\bibinfo  {journal} {Journal of Sol-Gel Science and Technology}\ }\textbf
  {\bibinfo {volume} {16}},\ \bibinfo {pages} {243} (\bibinfo {year}
  {1999})}\BibitemShut {NoStop}%
\bibitem [{\citenamefont {George}\ and\ \citenamefont
  {Seena}(2012)}]{Geor:2012}%
  \BibitemOpen
  \bibfield  {author} {\bibinfo {author} {\bibfnamefont {A.}~\bibnamefont
  {George}}\ and\ \bibinfo {author} {\bibfnamefont {P.~T.}\ \bibnamefont
  {Seena}},\ }\href {\doibase 10.1007/s10973-011-2042-3} {\bibfield  {journal}
  {\bibinfo  {journal} {Journal of Thermal Analysis and Calorimetry}\ }\textbf
  {\bibinfo {volume} {110}},\ \bibinfo {pages} {1037} (\bibinfo {year}
  {2012})}\BibitemShut {NoStop}%
\bibitem [{\citenamefont {Deville}(2008)}]{Devi:2008}%
  \BibitemOpen
  \bibfield  {author} {\bibinfo {author} {\bibfnamefont {S.}~\bibnamefont
  {Deville}},\ }\href {http://dx.doi.org/10.1002/adem.200700270} {\bibfield
  {journal} {\bibinfo  {journal} {Adv. Eng. Mater.}\ }\textbf {\bibinfo
  {volume} {10}},\ \bibinfo {pages} {155} (\bibinfo {year} {2008})}\BibitemShut
  {NoStop}%
\bibitem [{\citenamefont {Deville}\ \emph {et~al.}(2011)\citenamefont
  {Deville}, \citenamefont {Viazzi}, \citenamefont {Leloup}, \citenamefont
  {Lasalle}, \citenamefont {Guizard}, \citenamefont {Maire}, \citenamefont
  {Adrien},\ and\ \citenamefont {Gremillard}}]{Devi:2011}%
  \BibitemOpen
  \bibfield  {author} {\bibinfo {author} {\bibfnamefont {S.}~\bibnamefont
  {Deville}}, \bibinfo {author} {\bibfnamefont {C.}~\bibnamefont {Viazzi}},
  \bibinfo {author} {\bibfnamefont {J.}~\bibnamefont {Leloup}}, \bibinfo
  {author} {\bibfnamefont {A.}~\bibnamefont {Lasalle}}, \bibinfo {author}
  {\bibfnamefont {C.}~\bibnamefont {Guizard}}, \bibinfo {author} {\bibfnamefont
  {E.}~\bibnamefont {Maire}}, \bibinfo {author} {\bibfnamefont
  {J.}~\bibnamefont {Adrien}}, \ and\ \bibinfo {author} {\bibfnamefont
  {L.}~\bibnamefont {Gremillard}},\ }\href {\doibase
  10.1371/journal.pone.0026474} {\bibfield  {journal} {\bibinfo  {journal}
  {PLoS ONE}\ }\textbf {\bibinfo {volume} {6}},\ \bibinfo {pages} {e26474}
  (\bibinfo {year} {2011})}\BibitemShut {NoStop}%
\bibitem [{\citenamefont {Duman}\ and\ \citenamefont
  {DeVries}(1974)}]{Duma:1974}%
  \BibitemOpen
  \bibfield  {author} {\bibinfo {author} {\bibfnamefont {J.~G.}\ \bibnamefont
  {Duman}}\ and\ \bibinfo {author} {\bibfnamefont {A.~L.}\ \bibnamefont
  {DeVries}},\ }\href {http://dx.doi.org/10.1038/247237a0} {\bibfield
  {journal} {\bibinfo  {journal} {Nature}\ }\textbf {\bibinfo {volume} {247}},\
  \bibinfo {pages} {237} (\bibinfo {year} {1974})}\BibitemShut {NoStop}%
\bibitem [{\citenamefont {Knight}\ and\ \citenamefont
  {Duman}(1986)}]{Knig:1986}%
  \BibitemOpen
  \bibfield  {author} {\bibinfo {author} {\bibfnamefont {C.~A.}\ \bibnamefont
  {Knight}}\ and\ \bibinfo {author} {\bibfnamefont {J.~G.}\ \bibnamefont
  {Duman}},\ }\href {\doibase http://dx.doi.org/10.1016/0011-2240(86)90051-9}
  {\bibfield  {journal} {\bibinfo  {journal} {Cryobiology}\ }\textbf {\bibinfo
  {volume} {23}},\ \bibinfo {pages} {256 } (\bibinfo {year}
  {1986})}\BibitemShut {NoStop}%
\bibitem [{\citenamefont {Griffith}\ \emph {et~al.}(1997)\citenamefont
  {Griffith}, \citenamefont {Antikainen}, \citenamefont {Hon}, \citenamefont
  {Pihakaski-Maunsbach}, \citenamefont {Yu}, \citenamefont {Chun},\ and\
  \citenamefont {Yang}}]{Grif:1997}%
  \BibitemOpen
  \bibfield  {author} {\bibinfo {author} {\bibfnamefont {M.}~\bibnamefont
  {Griffith}}, \bibinfo {author} {\bibfnamefont {M.}~\bibnamefont
  {Antikainen}}, \bibinfo {author} {\bibfnamefont {W.-C.}\ \bibnamefont {Hon}},
  \bibinfo {author} {\bibfnamefont {K.}~\bibnamefont {Pihakaski-Maunsbach}},
  \bibinfo {author} {\bibfnamefont {X.-M.}\ \bibnamefont {Yu}}, \bibinfo
  {author} {\bibfnamefont {J.~U.}\ \bibnamefont {Chun}}, \ and\ \bibinfo
  {author} {\bibfnamefont {D.~S.~C.}\ \bibnamefont {Yang}},\ }\href {\doibase
  10.1111/j.1399-3054.1997.tb04790.x} {\bibfield  {journal} {\bibinfo
  {journal} {Physiologia Plantarum}\ }\textbf {\bibinfo {volume} {100}},\
  \bibinfo {pages} {327} (\bibinfo {year} {1997})}\BibitemShut {NoStop}%
\bibitem [{\citenamefont {Mizrahy}\ \emph {et~al.}(2013)\citenamefont
  {Mizrahy}, \citenamefont {Bar-Dolev}, \citenamefont {Guy},\ and\
  \citenamefont {Braslavsky}}]{Mizr:2013}%
  \BibitemOpen
  \bibfield  {author} {\bibinfo {author} {\bibfnamefont {O.}~\bibnamefont
  {Mizrahy}}, \bibinfo {author} {\bibfnamefont {M.}~\bibnamefont {Bar-Dolev}},
  \bibinfo {author} {\bibfnamefont {S.}~\bibnamefont {Guy}}, \ and\ \bibinfo
  {author} {\bibfnamefont {I.}~\bibnamefont {Braslavsky}},\ }\href {\doibase
  10.1371/journal.pone.0059540} {\bibfield  {journal} {\bibinfo  {journal}
  {PLoS ONE}\ }\textbf {\bibinfo {volume} {8}},\ \bibinfo {pages} {e59540}
  (\bibinfo {year} {2013})}\BibitemShut {NoStop}%
\bibitem [{Note2()}]{Note2}%
  \BibitemOpen
  \bibinfo {note} {The ice-shaping properties of ZrAc are independent of the
  supplier.}\BibitemShut {Stop}%
\bibitem [{\citenamefont {Koga}\ \emph {et~al.}(2011)\citenamefont {Koga},
  \citenamefont {Westh}, \citenamefont {Nishikawa},\ and\ \citenamefont
  {Subramanian}}]{Koga:2011}%
  \BibitemOpen
  \bibfield  {author} {\bibinfo {author} {\bibfnamefont {Y.}~\bibnamefont
  {Koga}}, \bibinfo {author} {\bibfnamefont {P.}~\bibnamefont {Westh}},
  \bibinfo {author} {\bibfnamefont {K.}~\bibnamefont {Nishikawa}}, \ and\
  \bibinfo {author} {\bibfnamefont {S.}~\bibnamefont {Subramanian}},\ }\href
  {\doibase 10.1021/jp108347b} {\bibfield  {journal} {\bibinfo  {journal} {J.
  Phys. Chem. B}\ }\textbf {\bibinfo {volume} {115}},\ \bibinfo {pages} {2995}
  (\bibinfo {year} {2011})}\BibitemShut {NoStop}%
\bibitem [{\citenamefont {Nilsson}\ \emph {et~al.}(2016)\citenamefont
  {Nilsson}, \citenamefont {Alfredsson}, \citenamefont {Bowron},\ and\
  \citenamefont {Edler}}]{Nils:2016}%
  \BibitemOpen
  \bibfield  {author} {\bibinfo {author} {\bibfnamefont {E.~J.}\ \bibnamefont
  {Nilsson}}, \bibinfo {author} {\bibfnamefont {V.}~\bibnamefont {Alfredsson}},
  \bibinfo {author} {\bibfnamefont {D.~T.}\ \bibnamefont {Bowron}}, \ and\
  \bibinfo {author} {\bibfnamefont {K.~J.}\ \bibnamefont {Edler}},\ }\href
  {\doibase 10.1039/C6CP01389A} {\bibfield  {journal} {\bibinfo  {journal}
  {Phys. Chem. Chem. Phys.}\ }\textbf {\bibinfo {volume} {18}},\ \bibinfo
  {pages} {11193} (\bibinfo {year} {2016})}\BibitemShut {NoStop}%
\bibitem [{\citenamefont {Mootz}\ and\ \citenamefont
  {Seidel}(1990)}]{Moot:1990}%
  \BibitemOpen
  \bibfield  {author} {\bibinfo {author} {\bibfnamefont {D.}~\bibnamefont
  {Mootz}}\ and\ \bibinfo {author} {\bibfnamefont {R.}~\bibnamefont {Seidel}},\
  }\href {\doibase 10.1007/BF01131293} {\bibfield  {journal} {\bibinfo
  {journal} {Journal of inclusion phenomena and molecular recognition in
  chemistry}\ }\textbf {\bibinfo {volume} {8}},\ \bibinfo {pages} {139}
  (\bibinfo {year} {1990})}\BibitemShut {NoStop}%
\bibitem [{\citenamefont {Zhang}\ and\ \citenamefont
  {Cremer}(2006)}]{Zhan:2006}%
  \BibitemOpen
  \bibfield  {author} {\bibinfo {author} {\bibfnamefont {Y.}~\bibnamefont
  {Zhang}}\ and\ \bibinfo {author} {\bibfnamefont {P.~S.}\ \bibnamefont
  {Cremer}},\ }\href
  {http://www.sciencedirect.com/science/article/pii/S1367593106001517}
  {\bibfield  {journal} {\bibinfo  {journal} {Current Opinion in Chemical
  Biology}\ }\textbf {\bibinfo {volume} {10}},\ \bibinfo {pages} {658}
  (\bibinfo {year} {2006})}\BibitemShut {NoStop}%
\bibitem [{\citenamefont {Collins}\ and\ \citenamefont
  {Washabaugh}(2009)}]{Coll:2009}%
  \BibitemOpen
  \bibfield  {author} {\bibinfo {author} {\bibfnamefont {K.~D.}\ \bibnamefont
  {Collins}}\ and\ \bibinfo {author} {\bibfnamefont {M.~W.}\ \bibnamefont
  {Washabaugh}},\ }\href {\doibase 10.1017/S0033583500005369} {\bibfield
  {journal} {\bibinfo  {journal} {Quarterly Reviews of Biophysics}\ }\textbf
  {\bibinfo {volume} {18}},\ \bibinfo {pages} {323} (\bibinfo {year}
  {2009})}\BibitemShut {NoStop}%
\bibitem [{\citenamefont {Zavitsas}(2016)}]{Zavi:2016}%
  \BibitemOpen
  \bibfield  {author} {\bibinfo {author} {\bibfnamefont {A.~A.}\ \bibnamefont
  {Zavitsas}},\ }\href
  {http://www.sciencedirect.com/science/article/pii/S135902941630070X}
  {\bibfield  {journal} {\bibinfo  {journal} {Current Opinion in Colloid \&
  Interface Science}\ }\textbf {\bibinfo {volume} {23}},\ \bibinfo {pages} {72}
  (\bibinfo {year} {2016})}\BibitemShut {NoStop}%
\bibitem [{\citenamefont {Lo~Nostro}\ and\ \citenamefont
  {Ninham}(2012)}]{LoN:2012}%
  \BibitemOpen
  \bibfield  {author} {\bibinfo {author} {\bibfnamefont {P.}~\bibnamefont
  {Lo~Nostro}}\ and\ \bibinfo {author} {\bibfnamefont {B.~W.}\ \bibnamefont
  {Ninham}},\ }\href {\doibase 10.1021/cr200271j} {\bibfield  {journal}
  {\bibinfo  {journal} {Chem. Rev.}\ }\textbf {\bibinfo {volume} {112}},\
  \bibinfo {pages} {2286} (\bibinfo {year} {2012})}\BibitemShut {NoStop}%
\bibitem [{\citenamefont {Mak}(1968)}]{Mak:1968}%
  \BibitemOpen
  \bibfield  {author} {\bibinfo {author} {\bibfnamefont {T.~C.~W.}\
  \bibnamefont {Mak}},\ }\href {\doibase 10.1139/v68-579} {\bibfield  {journal}
  {\bibinfo  {journal} {Can. J. Chem.}\ }\textbf {\bibinfo {volume} {46}},\
  \bibinfo {pages} {3491} (\bibinfo {year} {1968})}\BibitemShut {NoStop}%
\bibitem [{\citenamefont {Clearfield}(1990)}]{Clea:1990}%
  \BibitemOpen
  \bibfield  {author} {\bibinfo {author} {\bibfnamefont {A.}~\bibnamefont
  {Clearfield}},\ }\href@noop {} {\bibfield  {journal} {\bibinfo  {journal}
  {Journal of Materials Research}\ }\textbf {\bibinfo {volume} {5}},\ \bibinfo
  {pages} {161} (\bibinfo {year} {1990})}\BibitemShut {NoStop}%
\bibitem [{\citenamefont {Hagfeldt}, \citenamefont {Kessler},\ and\
  \citenamefont {Persson}(2004)}]{Hagf:2004}%
  \BibitemOpen
  \bibfield  {author} {\bibinfo {author} {\bibfnamefont {C.}~\bibnamefont
  {Hagfeldt}}, \bibinfo {author} {\bibfnamefont {V.}~\bibnamefont {Kessler}}, \
  and\ \bibinfo {author} {\bibfnamefont {I.}~\bibnamefont {Persson}},\ }\href
  {http://dx.doi.org/10.1039/B402804J} {\bibfield  {journal} {\bibinfo
  {journal} {Dalton Trans.}\ ,\ \bibinfo {pages} {2142}} (\bibinfo {year}
  {2004})}\BibitemShut {NoStop}%
\bibitem [{\citenamefont {Rao}\ \emph {et~al.}(2007)\citenamefont {Rao},
  \citenamefont {Holerca}, \citenamefont {Klein},\ and\ \citenamefont
  {Pophristic}}]{Rao:2007}%
  \BibitemOpen
  \bibfield  {author} {\bibinfo {author} {\bibfnamefont {N.}~\bibnamefont
  {Rao}}, \bibinfo {author} {\bibfnamefont {M.~N.}\ \bibnamefont {Holerca}},
  \bibinfo {author} {\bibfnamefont {M.~L.}\ \bibnamefont {Klein}}, \ and\
  \bibinfo {author} {\bibfnamefont {V.}~\bibnamefont {Pophristic}},\ }\href
  {\doibase 10.1021/jp0734880} {\bibfield  {journal} {\bibinfo  {journal} {J.
  Phys. Chem. A}\ }\textbf {\bibinfo {volume} {111}},\ \bibinfo {pages} {11395}
  (\bibinfo {year} {2007})}\BibitemShut {NoStop}%
\bibitem [{\citenamefont {Singhal}\ \emph {et~al.}(1996)\citenamefont
  {Singhal}, \citenamefont {Toth}, \citenamefont {Lin},\ and\ \citenamefont
  {Affholter}}]{Sing:1996}%
  \BibitemOpen
  \bibfield  {author} {\bibinfo {author} {\bibfnamefont {A.}~\bibnamefont
  {Singhal}}, \bibinfo {author} {\bibfnamefont {L.~M.}\ \bibnamefont {Toth}},
  \bibinfo {author} {\bibfnamefont {J.~S.}\ \bibnamefont {Lin}}, \ and\
  \bibinfo {author} {\bibfnamefont {K.}~\bibnamefont {Affholter}},\ }\href
  {\doibase 10.1021/ja9602399} {\bibfield  {journal} {\bibinfo  {journal} {J.
  Am. Chem. Soc.}\ }\textbf {\bibinfo {volume} {118}},\ \bibinfo {pages}
  {11529} (\bibinfo {year} {1996})}\BibitemShut {NoStop}%
\bibitem [{\citenamefont {Gossard}\ \emph {et~al.}(2014)\citenamefont
  {Gossard}, \citenamefont {Toquer}, \citenamefont {Grandjean},\ and\
  \citenamefont {Grandjean}}]{Goss:2014}%
  \BibitemOpen
  \bibfield  {author} {\bibinfo {author} {\bibfnamefont {A.}~\bibnamefont
  {Gossard}}, \bibinfo {author} {\bibfnamefont {G.}~\bibnamefont {Toquer}},
  \bibinfo {author} {\bibfnamefont {S.}~\bibnamefont {Grandjean}}, \ and\
  \bibinfo {author} {\bibfnamefont {A.}~\bibnamefont {Grandjean}},\ }\href
  {http://dx.doi.org/10.1007/s10971-014-3409-2} {\bibfield  {journal} {\bibinfo
   {journal} {Journal of Sol-Gel Science and Technology}\ }\textbf {\bibinfo
  {volume} {71}},\ \bibinfo {pages} {571} (\bibinfo {year} {2014})}\BibitemShut
  {NoStop}%
\bibitem [{\citenamefont {Clearfield}(1964)}]{Clea:1964}%
  \BibitemOpen
  \bibfield  {author} {\bibinfo {author} {\bibfnamefont {A.}~\bibnamefont
  {Clearfield}},\ }\href@noop {} {\bibfield  {journal} {\bibinfo  {journal}
  {Reviews of Pure and Applied Chemistry}\ }\textbf {\bibinfo {volume} {14}},\
  \bibinfo {pages} {91} (\bibinfo {year} {1964})}\BibitemShut {NoStop}%
\bibitem [{\citenamefont {Leinala}, \citenamefont {Davies},\ and\ \citenamefont
  {Jia}(2002)}]{Lein:2002}%
  \BibitemOpen
  \bibfield  {author} {\bibinfo {author} {\bibfnamefont {E.~K.}\ \bibnamefont
  {Leinala}}, \bibinfo {author} {\bibfnamefont {P.~L.}\ \bibnamefont {Davies}},
  \ and\ \bibinfo {author} {\bibfnamefont {Z.}~\bibnamefont {Jia}},\ }\href
  {\doibase http://dx.doi.org/10.1016/S0969-2126(02)00745-1} {\bibfield
  {journal} {\bibinfo  {journal} {Structure}\ }\textbf {\bibinfo {volume}
  {10}},\ \bibinfo {pages} {619 } (\bibinfo {year} {2002})}\BibitemShut
  {NoStop}%
\bibitem [{\citenamefont {Sun}\ \emph {et~al.}(2014)\citenamefont {Sun},
  \citenamefont {Lin}, \citenamefont {Campbell}, \citenamefont {Allingham},\
  and\ \citenamefont {Davies}}]{Sun:2014}%
  \BibitemOpen
  \bibfield  {author} {\bibinfo {author} {\bibfnamefont {T.}~\bibnamefont
  {Sun}}, \bibinfo {author} {\bibfnamefont {F.-H.}\ \bibnamefont {Lin}},
  \bibinfo {author} {\bibfnamefont {R.~L.}\ \bibnamefont {Campbell}}, \bibinfo
  {author} {\bibfnamefont {J.~S.}\ \bibnamefont {Allingham}}, \ and\ \bibinfo
  {author} {\bibfnamefont {P.~L.}\ \bibnamefont {Davies}},\ }\href
  {http://www.sciencemag.org/content/343/6172/795.abstract} {\bibfield
  {journal} {\bibinfo  {journal} {Science}\ }\textbf {\bibinfo {volume}
  {343}},\ \bibinfo {pages} {795} (\bibinfo {year} {2014})}\BibitemShut
  {NoStop}%
\bibitem [{\citenamefont {Koga}, \citenamefont {Sebe},\ and\ \citenamefont
  {Nishikawa}(2013)}]{Koga:2013}%
  \BibitemOpen
  \bibfield  {author} {\bibinfo {author} {\bibfnamefont {Y.}~\bibnamefont
  {Koga}}, \bibinfo {author} {\bibfnamefont {F.}~\bibnamefont {Sebe}}, \ and\
  \bibinfo {author} {\bibfnamefont {K.}~\bibnamefont {Nishikawa}},\ }\href
  {\doibase 10.1021/jp3082744} {\bibfield  {journal} {\bibinfo  {journal} {J.
  Phys. Chem. B}\ }\textbf {\bibinfo {volume} {117}},\ \bibinfo {pages} {877}
  (\bibinfo {year} {2013})}\BibitemShut {NoStop}%
\end{thebibliography}%
\end{document}